\documentclass[aps,prb,amsmath,showpacs,superscriptaddress,twocolumn,reprint]{revtex4}
\usepackage{bm}
\usepackage{graphicx}
\usepackage{subfigure}
\DeclareMathOperator{\Real}{Re}
\DeclareMathOperator{\Imag}{Im}

\begin{document}
\title{Coherent spectrum rearrangement in graphene: extra nodal point 
and impurity band with anomalous dispersion}
 
\author{Yuriy~V.~Skrypnyk}
\affiliation{G. V. Kurdyumov Institute of Metal Physics,\\ 
             National Academy of Sciences of Ukraine,
             Vernadsky Ave. 36,
             Kyiv 03680, Ukraine}
\author{Vadim~M.~Loktev}
\affiliation{Bogolyubov Institute for Theoretical Physics,\\
             National Academy of Sciences of Ukraine,
             Metrolohichna Str. 14-b,
             Kyiv 03680, Ukraine}
\affiliation{National Technical University of Ukraine "Kyiv Polytechnic Institute",\\ 
               Prospect Peremogy 37, 
               Kyiv 03056, Ukraine}


\begin{abstract}
It is demonstrated that an extra nodal point and a domain with anomalous 
dispersion are present in the spectrum of charge carriers in graphene, when 
the concentration of weakly bound impurities exceeds a certain critical value, 
determined by the spatial overlap of individual impurity states. Corresponding 
spectrum rearrangement is shown to be of the cross type.
\end{abstract}

\maketitle

\section{Introduction}

Chemical functionalization of graphene is currently a rapidly developing field of
research.\cite{gem,buh} By depositing various atoms, molecules and chemically 
active groups on graphene, it is possible to tune up its properties according to 
specific technical applications. Undoubtedly, transport properties of graphene 
are decisive for its future in microelectronic devices of the new generation. With 
respect to the transport properties, adatoms should be viewed, first of all, as a 
sort of defects, which can significantly alter the spectrum of charge carriers and 
lead to the localization of states. It was demonstrated experimentally that ion 
bombardment\cite{fuh}, deposition of atomic hydrogen\cite{eli} or 
fluorine\cite{sav} opens a mobility gap in the electron spectrum of graphene 
and triggers the metal-insulator transition, just as it happens in conventional 
systems with massive carriers. The opening of the mobility gap in graphene is
attributed to the presence of impurity resonance states in the Dirac point vicinity. 
Moreover, it became evident that some amount of resonant defects is always 
present in graphene obtained by the micromechanical cleavage process.\cite{gei} 
It follows from theoretical studies of generic models for adsorbed atoms that 
such defects should considerably modify the electron density of states and affect 
the conductivity of graphene.\cite{fal, weh, kat,net,ihn} However, only those cases 
were examined, in which impurity levels of adatoms were strongly bound to the 
host.

Lifshitz defects (vacancies including), which do nothing to the system but change 
effective potentials on lattice sites occupied by them, bear a lot of similarity to strongly 
bound impurity centers.\cite{mit} Albeit such defects are capable in producing resonance 
states in the Dirac point vicinity, the corresponding resonances are of a reduced integral 
intensity, and the subsequent spectrum rearrangement has a diffuse, anomalous 
character, which is inherent in low-dimensional systems.\cite{sl,psl} This type of 
the spectrum rearrangement is characterized by the presence of two non-overlapping 
dispersion branches in the spectrum, which are separated by a mobility gap. At that, the 
density of states inside this gap can considerably exceed the one in the host.\cite{psl} 

However, there is a different opportunity: spectra of elementary excitations in
tree-dimensional systems (and in low-dimensional ones as well) can undergo a 
rearrangement of the coherent (cross) type with increasing the amount of 
impurities.\cite{ilp} The renormalized dispersion resulting from the spectrum 
rearrangement of this type looks similar to the hybridization between the host branch 
and the dispersionless branch that corresponds to the impurity state energy. Namely, two 
overlapping branches develop in the spectrum of a disordered system. As a consequence, 
two different energies of quasiparticles correspond to each wave vector. These dispersion 
branches are separated by a true gap, which is gradually broadening with increasing the 
impurity concentration. The cross-type spectrum rearrangement is usually accompanied 
by strong single-impurity resonances, which appear in low-dimensional systems only for 
weakly bound impurities. Thus, we examine below possible changes in the electron spectrum 
of graphene, which can occur under an increase in the impurity concentration, when 
impurity levels are weakly bound to the host.

\section{Impurity model}

When dealing with $\pi$-electron bands of graphene within the tight-binding
approximation, it is usually sufficient to take into consideration only matrix elements 
between nearest neighbors in the honeycomb lattice:\cite{rmf}
\begin{equation}
{\bm H}={\bm H}_{0}+{\bm H}_{im},\quad {\bm H}_{0}=\frac{1}{\sqrt{\pi\sqrt{3}}}%
\sum\nolimits' c_{{\bm n}\alpha}^{\dagger}c_{{\bm m}\beta}^{\phantom{\dagger}},
\label{ham0}
\end{equation}
where  ${\bm H}_{0}$ is the dimensionless host Hamiltonian, ${\bm H}_{im}$ is the impurity 
perturbation, vector ${\bm n}$ runs over lattice cells, indices $\alpha$ and $\beta$ enumerate 
sublattices, $c_{{\bm n}\alpha}^{\dagger}$ and $c_{{\bm n}\alpha}^{\phantom{\dagger}}$
are the creation and annihilation Fermi operators at the corresponding lattice site, and the prime
at the sum sign indicates that $\bm{n}\alpha\ne\bm{m}\beta$. For the sake of simplicity we 
choose the energy unit in such a way that the transfer integral $t=1/\sqrt{\pi\sqrt{3}}$. Since 
its magnitude in graphene is taken to be around $2.7$ eV in most cases, the adopted energy 
unit is about $6.3$ eV.

If chemisorbed atoms play the role of defects,  even a minimal impurity model should 
contain, in a general case, a possibility for the electron transfer from the host to some energy 
level that belongs to the adsorbed atom. This task is fulfilled by the well-known Fano-Anderson 
impurity model,\cite{fan} which was adopted in quite a few studies devoted to adatoms on 
graphene.\cite{fal, weh, kat,net,gru} On the assumption that impurities are deposited without 
any spatial correlation, we arrive at the following impurity part of the Hamiltonian:
\begin{equation}
{\bm H}_{im}=\sum\eta_{{\bm n}\alpha}%
^{\phantom{\dagger}}\left[\varepsilon_{0}d_{{\bm n}\alpha}^{\dagger}%
d_{{\bm n}\alpha}^{\phantom{\dagger}}+(\gamma c_{{\bm n}\alpha}^{\dagger}%
d_{{\bm n}\alpha}^{\phantom{\dagger}}+h.c.)\right],
\label{hamsd}
\end{equation} 
where dimensionless $\gamma$ is the hybridization between the adatom and the host, 
$\varepsilon_{0}$ is the bare energy of the adatom level, $d_{{\bm n}\alpha}^{\dagger}$
and $d_{{\bm n}\alpha}^{\phantom{\dagger}}$ are the creation and annihilation 
operators at this level, the variable $\eta_{{\bm n}\alpha}$ takes the value of 1 with
the probability $c$ or the value of 0 with the probability $1-c$, and $c$ is the 
impurity concentration. 

By eliminating wavefunction amplitudes at adatoms from the stationary Shr\"odinger
equation for the operators (\ref{ham0}) and (\ref{hamsd}), the problem can be reduced to a 
more simple form with
\begin{equation}
{\bm H}_{im}=V\sum\limits_{{\bm n}\alpha}\eta_{{\bm n}\alpha}^{\phantom{\dagger}}
c_{{\bm n}\alpha}^{\dagger}c_{{\bm n}\alpha}^{\phantom{\dagger}},\quad
V=\frac{|\gamma|^{2}}{\varepsilon-\varepsilon_{0}}
\end{equation}
A conclusion on the presence of a resonance state for an isolated impurity can be made 
based on the single-site $T$-matrix,
\begin{equation}
\tau(\varepsilon)=\frac{V}{1-g_{0}(\varepsilon)V}=%
\frac{|\gamma|^{2}}{\varepsilon-\varepsilon_{0}-|\gamma|^{2}g_{0}(\varepsilon)},
\label{tm}
\end{equation} 
where $g_{0}(\varepsilon)$ is the diagonal element of the host Green's function, 
\begin{equation} 
{\bm g}=\frac{1}{\varepsilon-{\bm H}_{0}},
\end{equation}
taken in the site representation. The required diagonal element can be approximated in a 
vicinity of the Dirac point ($\varepsilon=0$) as follows: 
\begin{equation}
g_{0}(\varepsilon)\approx 2\varepsilon\ln(|\varepsilon|)-i\pi|\varepsilon|,\quad%
|\varepsilon|\ll 1.
\label{apr}
\end{equation}
Substituting this expression into Eq.~(\ref{tm}) and taking into account that logarithm 
is a slowly varying function, we have: 
\begin{equation}
\tau(\varepsilon)=\frac{|\gamma_{ef}|^{2}}{\varepsilon-\varepsilon_{ef}+%
i\pi|\gamma_{ef}|^{2}|\varepsilon|},
\label{tf}
\end{equation}
where the effective hybridization and the effective energy of the impurity level are 
introduced: 
\begin{gather}
|\gamma_{ef}|^{2}=\frac{|\gamma|^{2}}{D},\,%
\varepsilon_{ef}=\frac{\varepsilon_{0}}{D},\nonumber\\
D=1-2|\gamma|^{2}\left(\ln|\varepsilon_{r}|+1\right).
\label{rt}
\end{gather}
It immediately follows from Eq.~(\ref{tf}) that $\varepsilon_{ef}\equiv\varepsilon_{r}$,
i.e. corresponds to the energy of a possible single-impurity resonance.

This resonance can be considered as a well-defined, when its damping is significantly less 
than the energy interval between $\varepsilon_{r}$ and the closest van Hove singularity 
of the host spectrum. In the case under consideration, the role of such singularity is played 
by the Dirac point. Thus, the mentioned condition for the appearance of a well-defined 
resonance is reduced to the inequality:
\begin{equation}
\delta\equiv\pi|\gamma_{ef}|^{2}\ll 1, 
\label{cond}
\end{equation} 
where $\delta$ is the small parameter that will be used further on. It is evident, that  
the inequality (\ref{cond}) is automatically satisfied at $|\gamma|\ll 1$. Thus, a 
well-defined resonance state is always present in a system with Dirac-like dispersion, 
when the impurity level is weakly bound and the resonance energy is located close to 
the Dirac point. For a relatively large hybridization constant, $|\gamma|\gtrsim 1$, 
the inequality (\ref{cond}) takes the form:
\begin{equation}
\frac{\pi}{2\left|\ln|\varepsilon_{r}|+1\right|}\ll 1,
\label{resl}
\end{equation}
which can be satisfied only at a close proximity of the resonance energy to the Dirac 
point, and coincides with the corresponding condition for Lifshitz impurity centers.\cite{sl} 
One can expect that at $|\gamma|\gtrsim 1$ the spectrum rearrangement, for the most part, 
also proceeds according to the anomalous scenario, which already has been examined in detail 
for Lifshitz impurities.\cite{sl, psl} Therefore, we are focused on the opposite case of weakly
bound impurities ($|\gamma|\ll 1$) below. In other words, the hybridization parameter has 
to be smaller than the transfer integral magnitude in the host. 

\section{Coherent spectrum rearrangement}

At low impurity concentrations, $c\ll 1$, the spectrum rearrangement can be analyzed 
by means of the modified propagator method.\cite{lang} Within this approach, the
self-energy $\sigma(\varepsilon)$, which approximates the averaged over impurity
distributions Green's function of a disordered system
\begin{equation}
{\bm G}=\left<\frac{1}{\varepsilon-{\bm H}}\right>\approx\frac{1}{\varepsilon-%
\sigma(\varepsilon)-{\bm H}_{0}},
\label{grf}
\end{equation}
is determined self-consistently:
\begin{equation}
\sigma(\varepsilon)=\frac{c|\gamma|^{2}}{\varepsilon%
-\varepsilon_{0}-|\gamma|^{2}g_{0}[\varepsilon-\sigma(\varepsilon)]}.
\label{mp}
\end{equation}
We would like to remind that this method does not work satisfactorily in all spectral intervals 
of a disordered system.\cite{ilp} Due to increased scatterings on impurity clusters, concentration 
broadening areas are formed nearby singular points of the spectrum. Inside these areas the 
approximation (\ref{mp}) is not valid and states turn out to be localized. Thus, in order to locate
such areas and to determine their widths, it is often enough to employ the standard Ioffe-Regel 
criterion:\cite{iof}
\begin{equation}
|\tilde{\varepsilon}(\varepsilon)|\equiv|\varepsilon-\Real\sigma(\varepsilon)|\gg%
-\Imag\sigma(\varepsilon),
\label{ir}
\end{equation} 
where $\tilde{\varepsilon}(\varepsilon)$ is the renormalized energy. According to the
approximation (\ref{grf}), it specifies the renormalized dispersion relation in a disordered system 
inside those spectral domains, which are occupied by extended states:
\begin{equation}
\tilde{\varepsilon}(\varepsilon)=\varepsilon(\bm{k}),
\label{tvar}
\end{equation}
where $\varepsilon(\bm{k})$ is the host dispersion. It is well known that the Hamiltonian 
(\ref{ham0}) yields the linear dispersion for free charge carriers, when the wave vector is counted
from one of the two inequivalent Dirac points  in the Brillouin zone:
\begin{equation}
\varepsilon(\bm{k})\approx\pm v_{F}k, \qquad 
v_{F}=\frac{a}{2}\sqrt{\frac{\sqrt{3}}{\pi}},
\end{equation}
where $a$ is the honeycomb lattice constant. 

Let's assume that the spectrum rearrangement has occurred already.  After the rearrangement, the
spectrum of the disordered system should undergo a cardinal change. We are going to pay attention
only to states that remain extended despite the presence of disorder and to make several estimations
that should help to sketch up the overall structure of the rearranged spectrum. Due to the singular 
character of the impurity perturbation, the distortion of the spectrum is most pronounced around the 
resonance energy $\varepsilon_{r}$. 

\begin{figure}
\subfigure[$c<c_{*}$]
{
\includegraphics[width=0.22\textwidth]{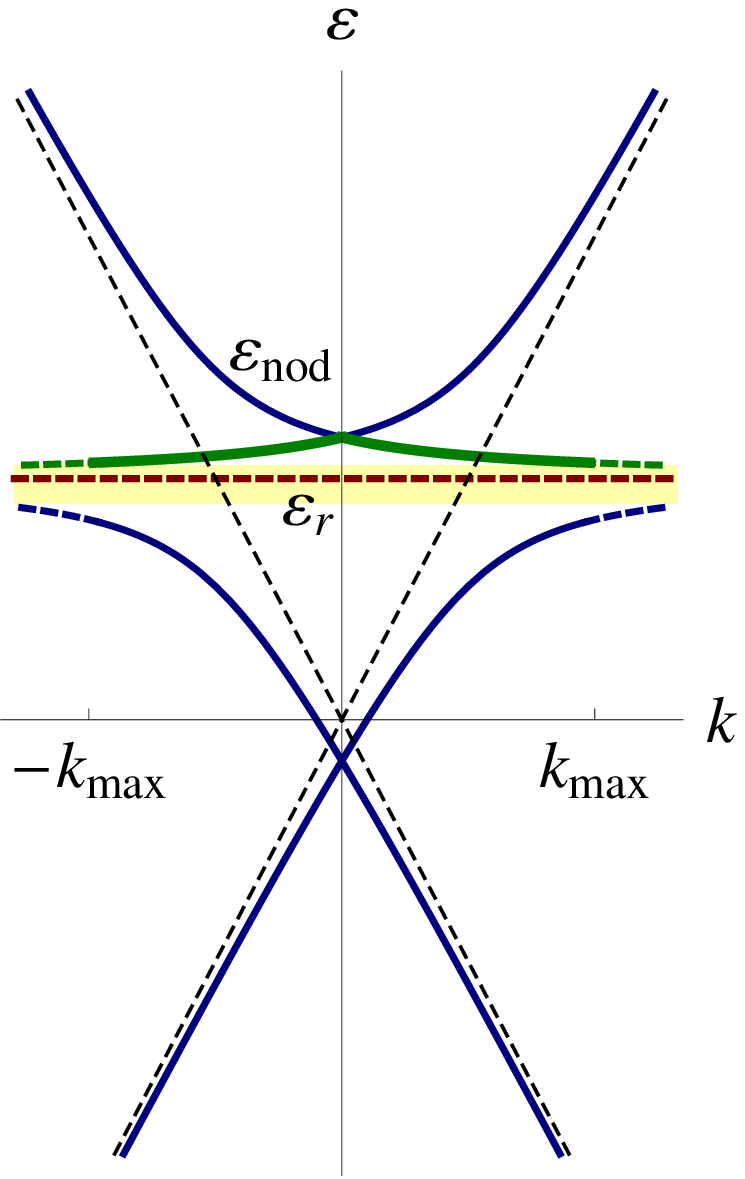}
\label{figa}
}
\subfigure[$c>c_{*}$]
{
\includegraphics[width=0.22\textwidth]{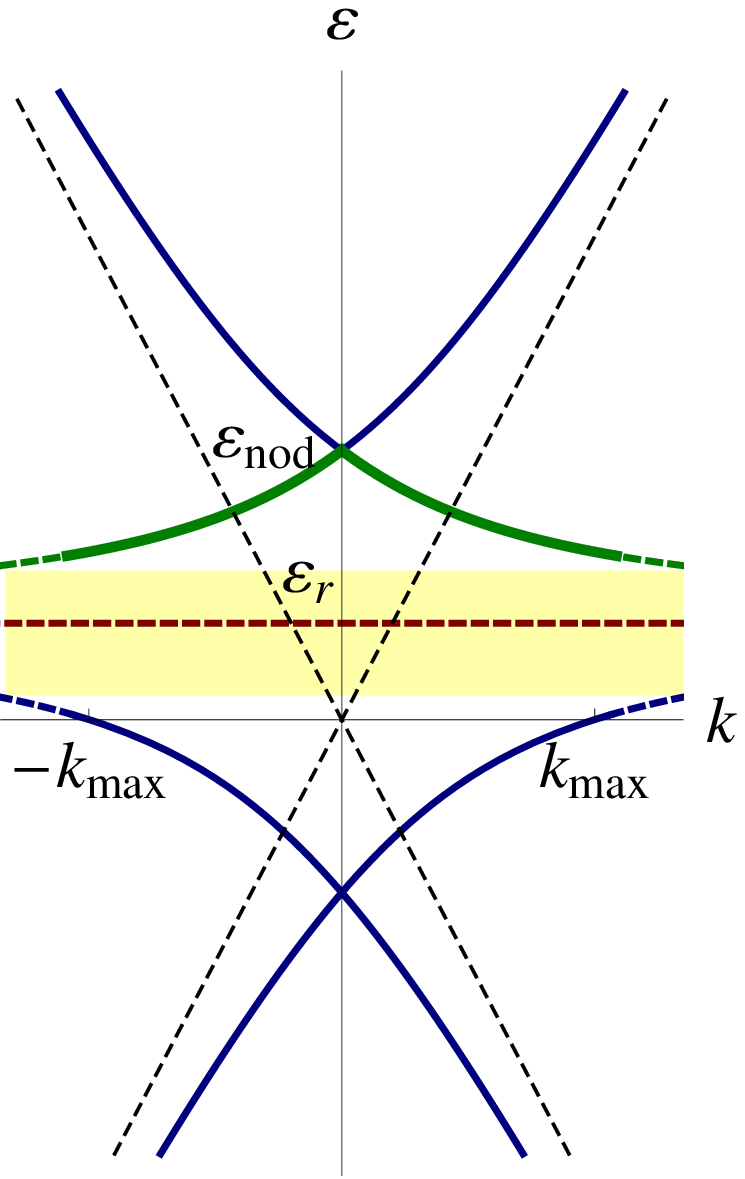}
\label{figb}
}
\caption{(Color online) Sketches of the dispersion of charge carriers in graphene with
weakly bound impurities after the cross-type spectrum rearrangement drawn according
to Eq.~(\ref{dl}): (a) $c_{csr}<c<c_{*}$, the Fermi velocity at the impurity nodal point
is significantly reduced; (b) $c_{*}<c\ll 1$, the Fermi velocity at the impurity nodal point
is comparable to the one in the host system. The host dispersion is shown by the dashed 
line. Branches with the anomalous dispersion are given by the thick (green) solid line. 
The concentration broadening area is marked by the (yellow) shade.}
\label{fig}
\end{figure}

Taking into account expressions (\ref{apr}-\ref{rt}), and (\ref{mp}), the imaginary part of the self-energy 
can be approximated as follows:
\begin{equation}
\Imag\sigma(\varepsilon)\approx-\frac{c\pi|\gamma_{ef}|^{4}%
|\tilde{\varepsilon}(\varepsilon)|}{(\varepsilon-\varepsilon_{r})^{2}+%
(\pi|\gamma_{ef}|^{2}\tilde{\varepsilon}(\varepsilon))^{2}},
\label{ims}
\end{equation}  
when the condition (\ref{ir}) holds, and, consequently, the corresponding states can be treated as extended 
ones. While remaining inside the domain of extended states, it is possible, regardless of the strong impurity 
scattering near the resonance energy, to go further and to omit the second term in the denominator of 
(\ref{ims}) (justification of this approximation will be provided below). Then, the inequality (\ref{ir}) in a 
vicinity of $\varepsilon_{r}$ can be rewritten as:
\begin{equation}
|\tilde{\varepsilon}(\varepsilon)|\gg\frac{c\pi|\gamma_{ef}|^{4}%
|\tilde{\varepsilon}(\varepsilon)|}{(\varepsilon-\varepsilon_{r})^{2}},
\end{equation}
which turns to
\begin{equation}
|\varepsilon-\varepsilon_{r}|\gg \sqrt{\pi c}|\gamma_{ef}|^{2}\equiv\Delta_{r}.
\label{brr}
\end{equation}
Thus, the width of the concentration broadening area of the impurity resonance $\Delta_{r}$ is 
increasing proportional to $\sqrt{c}$, i.e. behaves like the width of the mobility gap, which opens under 
the anomalous spectrum rearrangement in graphene with Lifshitz impurities.\cite{sl} However, it is worth 
noticing that, in comparison with the case of Lifshitz defects, the width $\Delta_{r}$ contains the small 
parameter of the problem $|\gamma_{ef}|^{2}$ as a multiplier. Similarly, within the same approximations, 
we have inside domains of extended states that
\begin{equation}
\tilde{\varepsilon}(\varepsilon)\approx\varepsilon-\frac{c|\gamma_{ef}|^{2}}%
{\varepsilon-\varepsilon_{r}}.
\label{tvara}
\end{equation}
The renormalized dispersion relation $\tilde{\varepsilon}(\bm{k})$, which is depicted in Fig.~\ref{fig}, 
can be obtained from Eqs.~(\ref{tvar}) and (\ref{tvara}) as a solution of the quadratic equation,
\begin{equation}
\tilde{\varepsilon}(\bm{k})\approx\frac{1}{2}\Bigl\{\varepsilon_{r}+\varepsilon(\bm{k}) \pm %
\sqrt{[\varepsilon_{r}-\varepsilon(\bm{k})]^{2}+4 c |\gamma_{ef}|^{2}}\ \Bigr\}.
\label{dl}
\end{equation}
Obviously, positions of nodal points, i.e. those points at which dispersion cones are touching each 
other by their vertices, are determined in the rearranged spectrum by the condition 
$\varepsilon(\bm{k})=0$. Thus, it follows from the dispersion law (\ref{dl}) that there should be 
a second (impurity) nodal point with the energy
\begin{equation}
\varepsilon_{nod}\approx\frac{1}{2}\Bigl(\varepsilon_{r}+\sqrt{\varepsilon_{r}^{2}+%
4 c |\gamma_{ef}|^{2}}\ \Bigr),
\label{enod}
\end{equation}
where we also assumed that $\varepsilon_{r}>0$ without loosing generality. At low impurity
concentrations,
\begin{equation}
\varepsilon_{nod}\approx \varepsilon_{r}+\Delta_{nod},\quad \Delta_{nod}\approx%
\frac{c |\gamma_{ef}|^{2}}{\varepsilon_{r}},\ 
\label{sgap}
\end{equation}
moreover, $\Delta_{nod}\ll \varepsilon_{r}$.
On the other hand, one can speak about the appearance of the \textit{extra} nodal point 
in the spectrum only in the case, when it lies outside the concentration broadening
area of the impurity resonance, i.e.
\begin{equation}
\Delta_{nod}\gg\Delta_{r}.
\label{ext}
\end{equation}  
In essence, the condition (\ref{ext}) determines the critical concentration of the
coherent spectrum rearrangement,
\begin{equation}
c_{csr}\sim\pi\varepsilon_{r}^{2}.
\label{csr}
\end{equation}
It is not difficult to see that this magnitude agrees with quick-and-dirty estimates for the 
critical concentration, which are based on the mutual spatial overlap of individual 
impurity states. Namely, the period of spatial oscillations of the host Green's function 
in graphene determines the characteristic radius of the single-impurity state: 
$r_{imp}\sim a\varepsilon_{r}^{-1}$. With increasing the impurity concentration, the 
average distance between impurities $\bar{r}\sim a c^{-1/2}$ is gradually decreasing.
Both lengths become equal at $c\sim\varepsilon_{r}^{2}$, which is in accord with 
the relation (\ref{csr}).\cite{psl,mit}

As it follows from Eqs.~(\ref{brr}) and (\ref{tvara}), the absolute value of the renormalized 
energy $\tilde{\varepsilon}(\varepsilon)$ is of the order of $\sqrt{c/\pi}$ at the boundaries of the
concentration broadening area located at the resonance energy, and, according to the relation 
(\ref{csr}), far exceeds $\varepsilon_{r}$ in the rearranged spectrum. The corresponding 
maximum value of the wave vector $k_{max}$ (see Fig.~\ref{fig}) that is reached at the 
boundaries of the concentration broadening area appears to be of the order of the inverse 
mean distance between impurities, as it would be expected:
\begin{equation}
k_{max}\sim\frac{2}{a}\sqrt{\frac{c}{\sqrt{3}}}.
\end{equation} 
By substituting the magnitude
$\sqrt{c/\pi}$ in place of $\tilde{\varepsilon}(\varepsilon)$ in the denominator of Eq.~(\ref{ims}), 
one can easily verify that the second term is small in comparison with the first one by virtue of 
the inequality (\ref{brr}). This justifies the omission of the second term in the denominator of 
Eq.~(\ref{ims}) for the extended states. 

When energy is crossing $\varepsilon_{r}$, the corresponding 
$\tilde{\varepsilon}(\varepsilon)\approx c|\gamma_{ef}|^{2}/(\varepsilon-\varepsilon_{r})$ 
changes its sign. In this way a spectral domain with anomalous dispersion develops between 
$\varepsilon_{r}$ and $\varepsilon_{nod}$ (see Fig.~\ref{figa}). With respect to the 
valence band (as it appears in the Figure), this domain can be considered as an 
\textit{impurity band} filled with extended (itinerant) states. Because of the particle-hole 
symmetry of the host spectrum, this impurity band fits exactly inside the gap that opens 
under the cross-type spectrum rearrangement of the conduction band. At that, the impurity 
band formation occurs simultaneously for both Dirac cones in the Brillouin zone of 
graphene. 

It follows from Eq.~(\ref{enod}) and the definition (\ref{sgap}) that at 
\begin{equation}
c\gg c_{*}\sim\frac{\varepsilon_{r}^{2}}{|\gamma_{ef}|^{2}}
\end{equation}
the width of the anomalous dispersion domain $\Delta_{nod}$ becomes of the order
of $\sqrt{c}|\gamma_{ef}|$, and thus scales differently with the impurity concentration.  
Since $c_{csr}/c_{*}\sim\delta$, the small parameter, which was defined by Eq.~(\ref{cond}), 
this characteristic concentration always exceeds the critical concentration of the spectrum 
rearrangement. At the onset of the spectrum rearrangement, the quantity $\Delta_{r}$ 
has a square root dependence on $c$, while $\Delta_{nod}$ has a linear one, which 
ensures its leading growth. By contrast, both widths have a square root dependence on 
the impurity concentration at $c>c_{*}$, while their ratio remains nearly constant:
\begin{equation}
\frac{\Delta_{r}}{\Delta_{nod}}\sim\sqrt{\delta},\quad c>c_{*}.
\end{equation} 
In addition, the width of the spectral domain with anomalous dispersion $\Delta_{nod}$
exceeds $|\varepsilon_{r}|$ at $c>c_{*}$ (see Fig.~\ref{figb}). 

The Fermi velocity $\tilde{v}_{F}$ at the extra nodal point $\varepsilon_{nod}$ can be 
expressed through the Fermi velocity of the host $v_{F}$ by means of Eq.~(\ref{dl}): 
\begin{multline}
\frac{\tilde{v}_{F}}{v_{F}}=\frac{1}{v_{F}}%
\left.\frac{d\,\tilde{\varepsilon}(\bm{k})}{d\, k}\right|_{k=0}=%
\left.\frac{d\,\tilde{\varepsilon}(\bm{k})}{d\,\varepsilon(\bm{k})}\right|_{\varepsilon(\bm{k})=0}%
\approx\\
\frac{1}{2}\left[1-\left(1+4\frac{c}{c_{*}}\right)^{-1/2}\right],
\label{fv}
\end{multline}
where the plus sign was chosen for the square root term in Eq.~(\ref{dl}). It is easy to see from
the ratio (\ref{fv}) that the Fermi velocity $\tilde{v}_{F}$ is reduced in $c/c_{*}$ times at 
$c\ll c_{*}$ in comparison with the one of the host. As a result, it can be difficult to observe the 
branch with the anomalous dispersion in real experiments. However, this velocity is gradually 
approaching $v_{F}/2$ for $c\gg c_{*}$, which eliminates indicated difficulties. Because the 
Fermi velocity near $\varepsilon_{nod}$ varies in several times, and the width of the
anomalous dispersion domain varies with the impurity concentration according to a different law, 
the characteristic concentration $c_{*}$ separates two dissimilar regimes of the coherent spectrum 
rearrangement.  

The width $\Delta_{r}$ is of the order of $\delta|\varepsilon_{r}|$ at $c\sim c_{csr}$, i.e. is 
substantially smaller than the interval between $\varepsilon_{r}$ and the initial position of the host 
Dirac point. This relationship becomes reversed at
\begin{equation}
c>c_{sym}\sim\frac{\varepsilon_{r}^{2}}{\pi|\gamma_{ef}|^{4}}\sim\frac{c_{*}}{\delta}%
\sim\frac{c_{csr}}{\delta^{2}}.
\end{equation}
Therefore, the asymmetry, which is present in the spectrum due to the nonzero energy of the
impurity resonance state, is effectively smeared out, when the impurity concentration is that high.
   
In the first approximation, as it follows from Eqs.~(\ref{mp}) and (\ref{sgap}), the impurities under
consideration behave in a close vicinity of $\varepsilon_{nod}$ at $c_{csr}<c<c_{*}$ like Lifshitz 
defects with the on-site impurity perturbation $v_{L}\approx \varepsilon_{r}/c$. 
According to the results obtained in Ref.~\onlinecite{sl}, the critical concentration of the 
anomalous spectrum rearrangement for such defects can be roughly estimated as 
$c_{asr}\sim 1/v_{L}^{2}$. Thus, the condition $c<c_{asr}\sim c^{2}/\varepsilon_{r}^{2}$ 
should be satisfied for the system to remain in the unrearranged regime. This condition reduces 
to $c>c_{csr}$, which is, for certain, fulfilled. It was demonstrated in Ref.~\onlinecite{sl} that in 
such a case the concentration broadening area of the Dirac point is exponentially small as compared 
to the bandwidth. Thus, its presence can be, to a known extent, overlooked.

Similarly, the effective Lifshitz impurity perturbation in the vicinity of $\varepsilon_{nod}$
is $v_{L}\sim|\gamma_{ef}|/\sqrt{c}$ at $c>c_{*}$, and the exponential smallness of the 
corresponding concentration broadening area is guaranteed by the inequality 
$c<c/|\gamma_{ef}|^{2}$, which is fulfilled on account of the condition (\ref{cond}).

The host Dirac point, which is always present in the spectrum, is also shifted
from its initial position due to the effect of impurities. As it directly follows
from Eq.~(\ref{dl}), the magnitude of this shift is about 
$- c |\gamma_{ef}|^{2}/|\varepsilon_{r}|$ at low impurity concentrations, which agrees with 
the width of the anomalous dispersion domain at $c>c_{csr}$. At that, this
shift occurs in a direction opposite to the concentration displacement of $\varepsilon_{nod}$
from $\varepsilon_{r}$.

For $c<c_{*}$, the effective magnitude of the impurity perturbation 
$v_{L}\sim -|\gamma_{ef}|^{2}/|\varepsilon_{r}|$ in a vicinity of the shifted host Dirac 
point, and the corresponding critical concentration $c_{asr}\sim\varepsilon^{2}_{r}%
/|\gamma_{ef}|^{4}$. The inequality $c_{*}<c_{asr}$ is again met in view of the 
condition (\ref{cond}). Thus, the width of the concentration broadening area remains 
once more exponentially small. At $c>c_{*}$, the Dirac point shift is 
$\sim\sqrt{c}|\gamma_{ef}|$ by the absolute value, and the analysis of its concentration
broadening can be performed in much the same way as it was done above for the extra 
nodal point vicinity. 

\section{Conclusion}

To summarize, with increasing the concentration of weakly bound impurity
centers, which yield well-defined resonance states, the electron spectrum of
graphene undergoes the coherent cross-type spectrum rearrangement. In
contrast to the already known cases of the coherent spectrum rearrangement,
it manifests the impurity nodal point in the spectrum and the impurity band
with the anomalous dispersion. In addition, it features a concentration broadening
area, or a mobility gap, around the impurity resonance energy, and thus, when
the Fermi level position in the system is controlled by, for example, the gate voltage, 
it should be possible to observe a metal-insulator transition on Fermi level's entering 
the mobility gap, and then a re-entrant insulator-metal transition on its leaving.
For Lifshitz defects, the mobility gap is narrower than the bandwidth in
$\sim\sqrt{c}$ times. In contrast, the mobility gap width for weakly bound defects 
$\Delta_{r}$ contains additional small parameter as a multiplier (see the inequality 
(\ref{brr})). Therefore, weakly bound impurities are more favorable for the 
experimental study of the re-entrant insulator-metal transition in graphene. 

\begin{acknowledgments}
This work was supported by SCOPES Grant $N^{o}$~IZ73Z0-128026 of the 
Swiss NSF, the SIMTECH Grant $N^{o}$~246937 of the European FP7 program, the 
State Program �Nanotechnologies and Nanomaterials�, Project No. 1.1.1.3, and by 
the Program for Fundamental Research of the Department of Physics and Astronomy 
of the NAS of Ukraine.
\end{acknowledgments}

\end{document}